\pdfoutput=1
\documentclass[12 pt]{article}
\usepackage[utf8]{inputenc}
\usepackage{authblk}
\usepackage{epigraph}
\usepackage{hyperref}
\usepackage{graphicx}
\usepackage{mathtools}   
\usepackage{amsfonts}
\usepackage{appendix}
\usepackage{caption}
\usepackage{natbib}
\usepackage{orcidlink}
\graphicspath{ {./figures/} }
\hypersetup{
    colorlinks=true,
    linkcolor=blue,
    urlcolor=blue,
}

\newtheorem{definition}{Definition}

\title{The Limitations of the Notion of `Observable' in Diffeomorphism-Invariant Models\thanks{Accepted for publication at \textit{The British Journal for Philosophy of Science}. Preprint of May 2026. Please cite published version when available.}}

\author[]{Álvaro Mozota Frauca}
\affil[ ]{alvaro.mozota@upc.edu, \orcidlink{0000-0002-7715-0563} \href{https://orcid.org/0000-0002-7715-0563}{https://orcid.org/
0000-0002-7715-0563}}
\affil[]{Department of Architectural Technology, Division of Mathematics\\
Universitat Politècnica de Catalunya. Barcelona (Spain)}

\date{\today}

\begin{document}

\maketitle

\begin{abstract}
The application of the notion of `observable' from gauge theory to diffeomorphism-invariant theories---most relevantly to general relativity---has led to numerous conceptual and technical issues when interpreting classical theories with this symmetry and building quantum versions of them. In this article I distinguish between two senses of gauge transformation: local and global, and I argue that the notion of observable appears more naturally in the local sense of gauge transformation. Then, I argue that diffeomorphism invariance can be understood as a gauge symmetry only from a global point of view, and hence, that the concept of observable applies only in a restricted manner. This has the consequence that some popular claims in the literature, such as the claim that the physical content of diffeomorphism-invariant models is encoded in correlations, are unfounded.
\end{abstract}

\section{Introduction}

The concepts and tools from gauge theory are a key ingredient in our understanding of contemporary high-energy physics. In gauge theories, instead of directly representing physically meaningful quantities, one uses a mathematically richer structure, which allows representing a physical situation in multiple ways, and which has been shown to be very fruitful in providing successful and empirically adequate theories. Diffeomorphism-invariant theories share some features with gauge theories, and this has motivated establishing an analogy between both kinds of theory to extract conclusions about the interpretation of general relativity and the way it ought to be quantized.

One of the concepts from gauge theory that has been very relevant in the foundations of general relativity and quantum gravity literature is the concept of `observable'. Intuitively speaking, an observable corresponds to a physical quantity described by the theory that is a property of the system described by the theory and that one could go and measure. This distinguishes it from other quantities in a gauge theory that are mere artifacts of the representation with no correspondence in the real world. That is, while an observable would be describing a genuine physical quantity, some other quantities in our formalism change from representation to representation and have no physical meaning.

In gauge theories, one can give a technical definition of observable. For this, one needs to introduce the notion of gauge transformation: it is a transformation in the formal structures of the theory that transforms a representation of a physical situation to a different representation of the same physical situation. If a quantity is an observable, it will remain unchanged under gauge transformations.

Trouble begins when one tries to apply this idea to general relativity and other diffeomorphism-invariant theories. In these theories, spacetime points or events are represented by points in a manifold. This representation is not unique, as there is great freedom at the time of assigning spacetime points to points in the manifold. Indeed, starting with any given representation, one can build an equivalent one by smoothly mapping points in the manifold to other points in the manifold. This transformation is known as diffeomorphism, and it is the symmetry that many take to be the gauge symmetry of general relativity.

By applying the notions of gauge theory in general, and of observable in particular, to this symmetry, many authors in the literature have interpreted general relativity and attempted to quantize it. However, in this article I want to follow other authors who have argued that the analogy between gauge theories and general relativity is limited, and that many of the claims in the literature are misguided. Some of the articles where criticisms have been raised are \citep{Kuchar1991, Kuchar1992, Kuchar1993, Isham1993, barbour_constraints_2008, Pitts2014-PITCIH, Pitts_2017, MozotaFrauca2023, mozota_frauca_gps_2024, mozota_frauca_quantum_2025, mozota_frauca_does_2026, Thebault2012, Gryb2016,Thebault2012,gryb_time_2023, pitts_peter_2022}.



Key to my argument is the observation that one should distinguish between two notions of gauge transformation: global and local. A local gauge transformation refers to a change in the representation of a system at a moment of time (or point in spacetime), while a global gauge transformation refers to a change in the representation of a whole history of a system. In gauge theories and most of the uses of the term gauge transformation, it doesn't make a big difference to think in gauge transformations in one way or another, but for the case of diffeomorphisms this difference is crucial. Diffeomorphisms can be understood as gauge transformations only from the global perspective, and, therefore, one should be careful at the time of applying notions and techniques of gauge theory to theories like general relativity, as only a few of them will apply, and in a restricted manner.

The distinction between global and local gauge transformations and its relevance for this debate has been known for a long time and plays an important role in some recent works on the topic \citep{Pitts2014-PITCIH,Pitts_2017,pitts_peter_2022, Gryb2016,Thebault2012,gryb_time_2023}. The argument in this article therefore shares the spirit of these earlier works, however, it is novel in several aspects. 

First, it puts the focus on the notion of observable, offers a clear conceptual distinction between the intuitive sense of observable and the technical definition in phase space, and explains how the two differ in the case of diffeomorphism invariant models. The conceptual kind of argument presented in this article can therefore be seen as complementary to other more technical arguments in the literature\footnote{Of course, the works and authors cited here also give conceptual arguments, but here the focus will be mostly on this aspect.}. For instance, \citep{Pitts_2017} analyzes models which share part of the constraint structure of general relativity to reject the technical definition of observable and \citep{gryb_time_2023} provide an exhaustive analysis and classification of symmetries action principles from the perspective of mathematical physics to conclude that reparametrization invariance differs in deep aspects from gauge symmetries. In this sense, by focusing just on the conceptual aspects of the debate, my article offers an interesting and accessible perspective. Similarly, while this article won't be addressing the historical perspective of the debate, I refer the interested reader to \citep{pitts_peter_2022}.

Second, my article explicitly defines global observables, which are physical quantities or predictions of a theory that do not need to be constrained in a specific instant of time or point in spacetime. Introducing this distinction is useful because it allows making claims like that whether two spacetime points are timelike separated or not is a prediction with observable consequences, even if this prediction is not naturally encoded as a phase space function. Indeed, while in the previous literature the focus lies mostly on local observables, my argument makes it explicit that global observables, including spatiotemporal relations, are also a key ingredient of our physical theories. 


Third, my article offers a novel analysis of the frozen observable view of reparametrization invariant theories. I argue that the fact that correlations can be written as phase space functions which are constant in time is just a feature of any deterministic theory with no further implication for its interpretation. Furthermore, I argue that any spatiotemporal relation can also be encoded in the form of evolution-invariant phase space functions, which directly challenges the claim of the frozen observable view that only correlations are invariant and observable. On top of this, I argue that even if these quantities satisfy the technical definition of observable, they are not observable in the intuitive, local sense of the term. The reason for this is that these functions are highly complicated functions which lack a straightforward relation with the properties of a system. In other words, while it is true that knowing the state of a system at a time allows computing highly non-trivial quantities, that does not mean that one is able to observe them in any direct sense of the term.

My analysis of the frozen observable view should make clear that at most one could try to impose an invariance condition only from a global perspective that takes phase space points to represent full histories of a system. But once this is acknowledged, one needs to accept the standard interpretation of diffeomorphism invariant theories in which spacetime models contain both correlations and spatiotemporal relations and one can think of local field quantities at spacetime points as local observables with no conceptual issue.

\section{Global and local gauge transformations in gauge theories} \label{sect_gauge_symmetries}


\subsection{Local and global observables} \label{subsect_local_global}

From a formal point of view, I will distinguish between two classes of models. On the one hand, in spacetime theories, models are built by defining a spacetime manifold $\mathcal{M}$ and a series of fields $\phi(x^{\mu})$ at each point in the manifold. On the other hand, we have theories in which one simply describes how a series of degrees of freedom $q(t)$ evolve in time, which is represented by a one-dimensional manifold. Typically, we understand these theories also as spatiotemporal, as $q(t)$ usually represents things like the positions of bodies in space. When understood as functions, we take that $q(t)$ and $\phi(x^{\mu})$ represent a whole evolution of $q$ (in time) and of $\phi(x^{\mu})$ (in spacetime). When we look at a given instant or point (specified by a temporal coordinate $t_0$ or set of spacetime coordinates $x^\mu$), these functions give us values of $q$ or $\phi$ (and their derivatives) for this instant or point.

The mathematical functions in our models, $q(t)$ and $\phi(x^{\mu})$, are taken to represent physical quantities in the world, even if this relationship is sometimes complicated, indirect, or debated. As advanced in the introduction, I will take gauge theories to be theories in which there are different representations of the same physical situation which are perfectly equivalent. I'll refer the reader to \citep{Henneaux1994,Rothe2010} for standard textbooks on gauge and constrained systems and to \citep{Healey2007, weatherall_understanding_2016} for more conceptual discussions. A gauge transformation is a (mathematical) transformation between two equivalent representations. However, for the discussion in this article it is important to distinguish between two types of gauge transformations: local and global.

\begin{definition}
\textbf{Local gauge transformation}: a local gauge transformation is a transformation of $q(t)$ or $\phi(x^\mu)$ (and their derivatives) at a given instant or point (specified by $t_0$ or $x^\mu_0$) such that the physical entities or properties represented at this instant or point remain unchanged. That is, the physical state of affairs at the instant identified by $t_0$ is equivalently represented by the values of $q(t)$ and (possibly) its derivatives at $t_0$, or by the values of the transformed $q'(t_0)$ and its derivatives at $t_0$. For the spacetime case, the same applies for $\phi(x)$ at a spacetime point identified by $x^\mu_0$.

%
%
%
\end{definition}

\begin{definition}
\textbf{Global gauge transformation}: a global gauge transformation is a transformation of $q(t)$ or $\phi(x^\mu)$ in the whole manifold (temporal or spatiotemporal) which leaves unchanged the history or spatiotemporal configuration of physical entities or properties represented by the model. 
\end{definition}

In the definition of local gauge transformations I have allowed for the presence of derivatives for two reasons. First, it is standard in gauge theories that some physical quantities are encoded in derivatives of configuration space variables. Second, even outside of gauge theories, the standard understanding of the local state of a system includes derivatives. For instance, in an $N$-body problem the velocities of the bodies are usually considered part of the instantaneous state of the system, and in general relativity one can consider the curvature (which is built from derivatives) to be an observable quantity at a spacetime point. In this sense, my definition aligns with the standard usage. Note that it differs from the more restrictive, ultra-local notion that excludes derivatives, as well as from the broader notion that takes local observables to be defined on finite, though small, regions rather than at strictly infinitesimal points. While these alternative notions are relevant in certain contexts--such as in the interpretation of observables associated with Wilson loops--for the purposes of this article it is sufficient to use the definition of locality given above.

When the gauge transformations can be understood from a local point of view, they can also be understood from a global point of view. However, the implication in the opposite direction does not hold, or so I will argue in this article. That is, I will argue that if a theory provides different equivalent representations of the history of a system, this does not mean that at each spacetime point (or instant of time) there is an associated change in representation. In section \ref{sect_diffeo} I will argue that this is what happens in the case of theories with an explicit diffeomorphism invariance.

Let me clarify that my definition of local gauge transformation differs from the definition of `local symmetry' that is common in the literature. By `local symmetry' one often refers to transformations with coordinate-dependent parameters. While it is uncontroversial that diffeomorphisms match this definition, I will argue that they are not local gauge symmetries in the sense that they do not provide different representations of what happens at an instant of time. Let me also clarify that in the context of gauge symmetries the terms `local' and `non-local' are also used to refer to whether generators have finite or infinite power expansions around a point. This is a technical sense of these term that is not the one intended in this article.

Now we can introduce the concept of `observable'. Although it is closely related to the Hamiltonian formalism, it is a concept that can be formulated in a formalism-free manner and then adapted to the different formalisms in which one may express physical models. The concept of observable aims to tell apart real physical content from mere representational artifacts. By `observables' we therefore refer to the physical properties of systems described by gauge theories. As before, here it will be convenient to distinguish between local and global senses of the term.

\begin{definition}\label{definition_local_observable}
\textbf{Observable (local)}: A local observable is a physical quantity or property at an instant of time or point in spacetime. Local observables remain unchanged by local gauge transformations. 
\end{definition}

\begin{definition}\label{definition_global_observable}
\textbf{Observable (global)}: A global observable is a physical quantity or property that isn't necessarily restricted to an instant of time or spacetime event. Global observables remain unchanged by global gauge transformations. 
\end{definition}

All local observables are global observables, but the definition of global observable is more general. For instance, for a system of two bodies gravitating around each other, we could include the minimal distance between the two bodies or the time that elapses between two configurations. Formally, we would say that local observables are functions of the fields or variables at a time or spacetime point, i.e., $O(q(t))$ or $O(\phi(x^\mu))$. Global observables are more general, as we can include functionals of the fields or variables for the whole history of the system, i.e., $O[\phi]$ or $O[q]$. 

From the category of global observables, there is one group that deserves special mention. These observables correspond to temporal or spatiotemporal relations. For instance, in Newtonian models we can ask whether a configuration of a system happens in between two others, and in relativistic theories there are global observables telling us whether two points are timelike separated and the proper time along a geodesic joining them. Clearly, these are usually considered to be part of the content of our models, even if they are not local observables.

The most common use of the term `observable' in the literature is in the local sense. The reason for this is because it is used in the context of quantum mechanics for describing the properties of a system that could be measured at a time. However, in the foundations and philosophy of physics, `observables' are sometimes defined in a wider sense. For instance, Rovelli defines them as `the physical quantities that we can predict and measure in real experiments' \citep{rovelli_gps_2002} and Gryb and Thebàult as `empirical quantity that the theory predicts a value for' \citep{gryb_time_2023}. This kind of definition seems to be wide enough for including global observables. For instance, the proper time along a trajectory in spacetime is a global quantity that general relativity predicts.

The definitions just given make more emphasis on the empirical connotations that the term `observable' carries than the one I have used and will be using in this article. In this sense, I take observables to be physical quantities that a theory predicts or describes and won't be that concerned with the way in which one could measure or observe these quantities. My phrasing will be a bit more realist than the operationalist-leaning of some part of the physics literature, but my claims can be readily translatable and the conclusions should be independent of it. For instance, when I claim that spatiotemporal structures are global observables in the sense that there are non-local facts in the worlds described by the models captured by these structures, this can also be understood in more operational terms, such as the ways in which spatiotemporal relations in special relativity can be phrased in terms of clocks and light rays.

In section \ref{sect_consequences} I will discuss how the wide definition of `observable' in the foundations of general relativity and quantum gravity literature has lead to a problematic mixture of the global and local sense of observable. Moreover, introducing the global sense of observable, helps me making emphasis in that I take spatiotemporal structures to be a fundamental ingredient of our physical models. In this sense, I will argue that the analysis that only take local observables or correlations to be physically relevant lead to problematic interpretations of diffeomorphism-invariant models.

While some authors have distinguished between local and global gauge transformations or between the instantaneous state of a system and a whole history, the distinction between local and global observables and the argument that global observables are needed for understanding an important part of our models haven't been put forward before. In this sense, my argument goes one step further than some other works. For instance, while Pitts \citep{Pitts_2017, pitts_peter_2022} gives arguments for why diffeomorphisms are not local gauge transformations, at the time of proposing reforming the technical definition of observable, his proposal is concerned with local observables. As far as I see, his proposal is compatible with claiming that there are global observables. Similarly, Gryb and Th\'ebault \citep{gryb_time_2023} distinguish symmetries-at-an-instant from other symmetries, but then at the time of defining what they call chronoobservable structures, they focus mostly on phase space functions, which are associated with the local sense of observable. It is true that they speak about the chronoordinal and chronometric aspects of time in mechanical models, but then they relate them to these chronoobservable structures which end being strongly related to functions in phase space. These functions are what they call clock observables. From my perspective, there is no problem in saying that the ordinal and metric aspects of time or spacetime are global observables and not worrying about how the instantaneous state of a system or subsystem can be used for `measuring' time. The distinction between the spacetime structures and their measurable effects on dynamical systems is of course subtle and key for debates about the nature of spacetime. However, for the aims of this article it will suffice with claiming that (at least) some spacetime structures are observable in the sense of part of our models with empirical consequences. While Gryb and Th\'ebault seem more reluctant to directly include spatiotemporal structures in the chronoobservable category, I believe they would agree with the core of my argument in the next section and agree that some of what I call global observables should be preserved and that they aren't if diffeomorphisms are interpreted as global symmetries.



\subsection{Technical definition of observable on phase space} \label{subsect_technical}

Having introduced the intuitive notions of global and local observables, we can turn to the formalization of theories with gauge transformations. The concept of gauge transformation applies even before introducing the Lagrangian or Hamiltonian formalization of physical theories. For instance, in the 4-potential formalism of electromagnetism, instead of directly representing the electromagnetic field $F_{\mu\nu}$, the 4-potential $A_{\mu}$ is used for convenience. Different 4-potentials can represent the same electromagnetic field, and the transformations that map a 4-potential to an equivalent one are the gauge transformations of the theory. These transformations can be understood both from a local and a global point of view, and similarly, one can think that the electromagnetic field, both at a spacetime point and its whole history, is the observable of the theory.

Gauge theories have a peculiar formal structure regarding their dynamics. There is under-determination in their laws: given a set of initial conditions, the equations of motion of the theory do not pick up a unique solution. However, this is just a consequence of the multiplicity of equivalent representations. That is, all the solutions of the equations of motion for a given set of initial conditions turn out to be equivalent, i.e., to represent the same history for the system. For instance, in the case of electromagnetism, the equations of motion for $A_{\mu}$ do not completely fix it, but any solution represents the same $F_{\mu\nu}$.

The second peculiarity of gauge theories at this level is that some of their laws are not dynamical laws as we usually understand them, but constraint equations. From a technical perspective, this means that some of their equations of motion are not second-order in temporal derivatives, but are just relations between the variables or fields and their first-order derivatives. This implies that not any set of initial conditions for the variables or fields and their derivatives is acceptable; only those satisfying these equations are. In the case of electromagnetism, only $A_{\mu}$s satisfying Gauss constraint are acceptable initial conditions.

When we move to the Lagrangian and Hamiltonian formalisms, we find similar structures: we will find versions of gauge transformations adapted to each formalism, and we will find a dynamics showing under-determination and, possibly, constraints. In the Lagrangian formalism, this is represented by a Lagrangian with a symmetry (a gauge symmetry), which means that minimizing the action given an initial and final configuration does not uniquely determine one trajectory in configuration space. When moving to the Hamiltonian formalism, the symmetry of the Lagrangian means that the mapping between the tangent bundle of configuration space and phase space is not invertible. This entails that some of the momenta cannot be freely chosen, and hence that the dynamics is restricted to just a subspace of phase space, the constrained surface. Moreover, the Hamiltonian is not uniquely defined in the whole phase space, as it can depend on some arbitrary functions. Different choices for these functions produce different solutions to the equations of motion, which are gauge-equivalent.

More precisely, the quantities that define the constrained surface are the primary constraints $\phi_a(q,p)$, by means of the relation $\phi_a(q,p)=0$. The Hamiltonian is $H(q,p)=H_c(q,p)+v^a\phi_a(q,p)$, where $v^a$ are arbitrary functions and I am using the convention that repeated indices imply a summation. Here I am using $q,p$ to refer to generic phase space variables, which could be field variables. Moreover, the condition that each of the primary constraints is satisfied along the evolution may produce new constraints, which are known as secondary constraints and which correspond to the constraints in the original theory and in the Lagrangian formalism. For instance, in the case of electromagnetism, Gauss law appears as a secondary constraint.

One can show that there is an intimate relation between gauge transformations and constraints in this formalism. Indeed, any two gauge-related trajectories in phase space are related by transformations generated by linear combinations of the constraints, known as the gauge generators $G_a$. This means that for any gauge equivalent pair $q(t),p(t)$ and $q'(t),p'(t)$ there is a pair of functions $q(t,\alpha),p(t,\alpha)$ such that $q(t,0),p(t,0)=q(t),p(t)$ and $q(t,1),p(t,1)=q'(t),p'(t)$ and which satisfy the equations:
\begin{eqnarray} \label{eqn_gauge_transformation_1}
\frac{\partial q}{\partial \alpha}&= \{q,G\} \\
 \label{eqn_gauge_transformation_2}
\frac{\partial p}{\partial \alpha}&= \{p,G\} \, ,
\end{eqnarray}
where the brackets $\{\cdot,\cdot\}$ represent the Poisson brackets in phase space and $G$ is the generator of the gauge transformation. Each gauge transformation has its own generator and they all differ in the coefficients in front of each of the constraints. These gauge transformations can be understood both from a local and a global point of view.

Given the form of gauge transformations \ref{eqn_gauge_transformation_1}-\ref{eqn_gauge_transformation_2}, it is straightforward to see that if a smooth function $O(q,p)$ on phase space satisfies the condition:
\begin{equation}\label{eqn_invariance}
 \{O,\Omega_i\}\approx0 \quad \forall \, \Omega_i \, ,
\end{equation}
where $\Omega_i$ represents all the constraints in the model and the symbol $\approx$ means that the equality is a weak equality, i.e., an equality just on the constraint surface, then $O(t)=O(q(t),p(t))$ remains invariant under gauge transformations. That is, if $q(t), p(t)$ and $q'(t),p'(t)$ are trajectories in the constrained surface related by a gauge transformation, $O(t)=O(q(t),p(t))=O(q'(t),p'(t))=O'(t)$\footnote{\label{footnote_defns_gauge}There are two competing technical definitions of gauge transformation in phase space. The technical definition of observable I will give here will work for both definitions. For discussions of these two approaches, see \citep{Pitts2013, Pitts2022, pitts_does_2024, Pooley2022,mozota_frauca_which_2024}}. This motivates the following technical definition of observable:

\begin{definition} \label{defn_observable_technical}
\textbf{Observable (phase space)}: An observable is a quantity in phase space that has weakly vanishing Poisson brackets with all the constraints in the model.\footnote{Let me clarify that this definition applies for the cases of interest of this article: standard gauge theories and diffeomorphism-invariant models. Models with second-class constraints require a slightly modified definition.}
\end{definition}

This technical definition of observable is better understood and motivated from the local point of view. Phase space is a space that encodes instantaneous configurations and velocities of a system and functions in phase space. Functions in phase space are therefore functions of the state of a system at a time. Observables, in this technical sense, correspond to observables in the local, intuitive sense (def. \ref{definition_local_observable}) for gauge theories. Again, electromagnetism is the paradigmatic case: observables in the technical case, i.e., invariant phase space functions, correspond to the electromagnetic field, i.e., the local observables of the theory.

However, let me mention here that despite the clear interpretation and motivation from the local point of view, in some discussions this definition is justified from the perspective of global gauge transformations. For now, let me just insist that, as phase space is a space of instantaneous states, the most natural interpretation of functions in this space is as local functions. I will come back to the possibility of a global justification for the technical definition in section \ref{sect_consequences}.


%

While this definition works well for the case of electromagnetism and gauge theories, it will have some important limitations for the case of diffeomorphism-invariant models. To the best of my knowledge, the impact of these limitations on the technical definition of observable have not been discussed in the literature before. Notice that the argument that leads to the technical definition of observable has the following structure. First, as $O$ has vanishing Poisson brackets with the constraints, then it is invariant. Second, as it is invariant, it is an observable in the intuitive sense. Let me turn to the problems with each of the pieces in turn.

A limitation of the second part of the argument is that it opens the door to mere mathematical quantities and quantities that are not directly observable in an intuitive sense. For instance, trivial constant functions in phase space, such as $O(q,p)=\pi$ ($\pi$ here represents the number), are observable in the technical sense. From an intuitive point of view, we cannot observe $\pi$ and it does not correspond to something described by theories like electromagnetism. Similarly, while it makes sense to claim that the electromagnetic field is observable in electromagnetism, the claim that a complicated function such as $\vec{E}^2+17\arctan(\vec{E}\cdot\vec{B})$ is an observable can be very reasonably challenged. Of course, it is a quantity we can compute if we know the values of the electromagnetic field, but it is not, or it does not seem to be, a property at a spacetime point. In this sense, one may want to add a condition that distinguishes intuitively observable quantities from quantities one may compute with them or mere constants. For theories like electromagnetism this point may sound like pedantry, but it will be relevant in the case of diffeomorphism-invariant theories. The reason for this is that some of the observables in the technical sense of this kind of model will clearly fall in the category of highly complicated phase space functions that have nothing to do with what an actual observer would observe when studying a system described by this kind of system.

A second limitation that plays a role in the case of diffeomorphism-invariant models is that the first part of the argument (if $O$ satisfies $\{O,\Omega_i\}\approx0$ $\forall$ $\Omega_i$, then it is invariant) does not hold in the opposite direction. That is, it is not necessary that all invariant quantities have vanishing Poisson brackets with the constraints. The reason for this is technical: one may have quantities that aren't defined in the whole phase space, or they may not be differentiable. This has even been acknowledged by defenders of the `gauge' view of diffeomorphism-invariant models \citep{dittrich_can_2017, Dittrich2015a, chataignier_relational_2024}, even if they haven't taken them to signal conceptual issues with this view. Given this non-differentiability, one cannot compute Poisson brackets for them. However, these quantities may be well-defined along the phase-space trajectories generated by gauge transformations \ref{eqn_gauge_transformation_1}-\ref{eqn_gauge_transformation_2} and they may be constant along these trajectories. This again seems like pedantry in the context of gauge theories, where we are used to continuous and differentiable functions, but it turns out that it will be very relevant in the case of theories like general relativity. I will give an example showing this in section \ref{sect_consequences}.

Finally, the crucial observation that needs to be made for the aims of this article is that the technical definition of observable in phase space takes the time coordinate $t$ to identify instants in time (or the spacetime coordinates $x^\mu$ to identify spacetime points). The technical definition of observables selects quantities $O(t)$ that are invariant, i.e., it selects quantities that in all possible representations assign the same values at a coordinate time $t$, which we use to identify moments in time. However, for the case of diffeomorphisms I will argue that this construction won't make sense, as what these transformations will do is to change the meaning of $t$. In this sense, I will argue that the formal definition of observable for gauge theories stops making sense in the case of theories like general relativity.

\section{Diffeomorphism invariance as a global symmetry}\label{sect_diffeo}

Now we can turn to diffeomorphism-invariant models and why they need to be treated and understood in quite a different way from gauge theories. As I have been discussing above, in our physical theories, we typically describe spacetime or time in terms of a manifold: a one-dimensional one for describing just time, or a higher-dimensional one (usually 4-dimensional) for describing spacetime. We describe physical variables or fields as functions (or other geometrical entities) on these manifolds. That is, we describe how variables evolve in time with functions $q(t)$ and how fields develop in spacetime with functions $\phi(x^\mu)$. We say that a physical theory is diffeomorphism-invariant when we interpret two diffeomorphism-related models of the theory to represent the same physical situation, i.e., the same configuration of fields in spacetime or the same temporal evolution of some variables.

I will define diffeomorphisms as a map from $x^\mu$ to $x'^\mu$, i.e., a smooth and invertible function $x'^\mu(x^{\mu})$. We require fields in the manifold to be transformed accordingly. For instance, if at a point $x^\mu$ the scalar field $\phi$ takes the value $\phi(x^\mu)$, the transformed field $\phi'$ will take the same value at the transformed point, i.e., $\phi'(x'^\mu)=\phi(x^\mu)$. Similarly, tensorial objects transform according to the tensor transformation rules. Notably, the metric tensor transforms as $g'_{\mu\nu}(x'^\mu)=g_{\alpha\beta}(x^\mu)\frac{\partial x^\alpha}{\partial x'^\mu}\frac{\partial x^\beta}{\partial x'^\nu}$. Diffeomorphisms can be interpreted as active or passive transformations, but for the discussion if this article this distinction won't be playing an important role. The important observation to be made is that the same coordinate point $x^\mu_0$ represents two different spacetime points before and after the diffeomorphism.

Now, from a purely formal point of view, any physical theory can be cast in a diffeomorphism-invariant manner. For this reason, it will be useful to introduce a couple of distinctions. We say that a theory is not explicitly diffeomorphism-invariant if it is expressed in terms of a preferred set of coordinates or class of coordinates. For instance, Newtonian mechanics is usually expressed in terms of Newton's absolute time $t$ and space $x$ or in terms of coordinates which follow an inertial reference system. Similarly, the laws of special relativistic theories are usually given in Lorentzian coordinates. 

Even if Newtonian mechanics and special relativity can be written in the language of differential geometry as theories on a manifold $M$ with some metric fields ($h_{\mu\nu}$ and $t_\mu$ in the Newtonian case, $\eta_{\mu\nu}$ in the special relativistic one), there is a further difference with theories like general relativity that motivates some authors not to consider them diffeomorphism-invariant. This difference is that even if one builds diffeomorphism-invariant versions of the model, they appear differently in the dynamics. For instance, in the manifold version of Newtonian or special relativistic theories the metric tensors $h_{\mu\nu}$ and $t_\mu$ or $\eta_{\mu\nu}$ do not have dynamical equations of their own, and they appear as fixed functions in the equations of motion for the degrees of freedom in the model. This is in contrast with general relativity, where the metric has equations of motion of its own, and these equations of motion are akin to those of gauge theories in that there is under-determination: given a set of initial conditions they only determine the metric (and any other field) up to a diffeomorphism. Let me say that in the latter case the theory has both dynamical and kinematical diffeomorphism invariance, while in the former case, it is only a kinematical invariance.

The issue of observables appears only in theories with explicit and dynamical diffeomorphism invariance. However, from my point of view, the interpretation of all these kinds of models is analogous regarding observables in the intuitive sense. In this article I will focus mostly in models with an explicit, dynamical diffeomorphism invariance and I will refer the interested reader to \citep{mozota_frauca_gps_2024} for a detailed defense of the claim that in all spacetime theories we understand observables in similar ways, independently of whether they are explicitly and dynamically diffomorphism invariant or not.

There are two models with an explicit and dynamical diffeomorphism invariance which have been widely discussed in the literature. One is, of course, general relativity. I will take models of general relativity to be given by a metric tensor $g_{\mu\nu}(x^{\mu})$ together with some matter degrees of freedom represented generically by $\phi(x^{\mu})$. Solutions of general relativity satisfy Einstein equations and the equations of motion of $\phi$. The other model is the Jacobi action of Newtonian systems. This model describes Newtonian systems but instead of using Newton's absolute time as a temporal coordinate, it uses an arbitrary time parameter and the models have an explicit and dynamical invariance under one-dimensional diffeomorphisms. In the discussion in this section I will focus just on general relativity, but I encourage the reader to take a look at some presentations of Jacobi models \citep{Barbour1994, gryb_jacobis_2010, MozotaFrauca2023, gryb_time_2023} to notice the strong resemblance between both kind of models.

In general relativity we use arbitrary coordinate systems and models related by arbitrary diffeomorphisms. Coordinates can be used to identify points in spacetime only in a given model. In a diffeomorphism-related version of a model, the same set of spacetime coordinates generally refers to different points. However, given one point in one model, we are able to identify the same point in some diffeomorphism-related model. There are several ways of doing so. If one knows which diffeomorphism relates the two models, it is straightforward that one needs to transform the coordinates of the point in one model according to the diffeomorphism in order to obtain the coordinates of the same point in the other model. For sufficiently rich models, one can use the configuration of fields or matter to identify points. For instance, if in a general relativistic model a configuration $\phi_i ^0$ only obtains at a spacetime point $x^\mu _0$, the same instant can be identified in a diffeomorphism-related version of the model by looking for the values of $x^\mu$ at which the configuration $\phi_i^0$ obtains. In other words, if $\phi_i^0$ only obtains once in a spacetime, $\phi_i(x^\mu), \phi'_i(x^\mu)$ are two diffeomorphism-related representations of such a spacetime, and $\phi_i(x^\mu _0)= \phi'_i(x'^\mu _0)=\phi_i ^0$, then $x^\mu_0$ and $x'^\mu_0$ represent the same point in spacetime in the model $\phi_i(x^\mu)$ and $\phi'_i(x^\mu)$, respectively.

Alternatively, one uses objects like clocks and rods to identify spacetime points or instants of time. The relationship $dt^2=-g_{\mu\nu}dx^\mu dx^\nu$ allows relating coordinates along curves in spacetime with the readings of clocks and rods. For instance, if we use two diffeomorphism-related models for describing the trajectory of a satellite launched from Earth with some given trajectory, we can use the readings of an ideal clock in this satellite to identify spacetime points. That is, two different diffeomorphism-related models will assign different spacetime coordinates to the spacetime point in which the clock reads 1 hour, but we are able to identify the point in both models because in each of the models, there is just one point that corresponds to the spacetime point 1 hour in the future from the point in which the clock reads 0.

Before getting into the technical details, we are already able to discuss what the observables from an intuitive point of view are. In the case of general relativity, clearly, any field at a spacetime point would count as a local observable. In our example, if the model includes a temperature field $T(x^\mu)$, it predicts the temperature along the trajectory of the rocket. Tensorial quantities are technically more involved, but one can convince oneself that they are also local quantities that the theory predicts. For instance, the electromagnetic field at a point is a local observable from this intuitive point of view. 


Global observables can be built from functionals of the local ones. Global observables include spatiotemporal relations, which in the case of general relativity are a quite rich familiy. For instance, in the example above we can include the proper time of the rocket along its trajectory in spacetime, and whether two spacetime points are time-like separated or not.


Given this intuitive understanding, we can turn to whether diffeomorphisms are gauge transformations. I take it to be uncontroversial that diffeomorphisms are global gauge transformations as I have defined them above. That is, diffeomorphisms are transformations of $q(t)$ or $\phi(x^{\mu})$ in the whole manifold that leave unchanged the history or spacetime configuration described by the models. 

The tricky aspect of diffeomorphisms and the definition of observable is in the local action of diffeomorphisms. Normally, the claim that diffeomorphisms are gauge transformations does not come supplemented with the discussion of whether these transformations are global or local. However, this makes a great difference, as diffeomorphisms are not local gauge transformations.


In the case of general relativity, if we transform the fields $\phi(x^\mu)$ into $\phi'(x^\mu)$ and look at a coordinate point $x^\mu _0$, what we are looking at is the configuration of the fields at two different spacetime points. In the example of the rocket, $x^\mu _0$ may represent the launch of the rocket in one model and the instant at which the clock reads 1 hour in the other. As the transformation does not relate two different representations of the state of affairs at the same spacetime point, it is not a local gauge transformation.

For any other diffeomorphism-invariant model, the diagnosis is the same. As after a diffeomorphism a coordinate $t_0$ or set of coordinates $x^\mu _0$ represents a different instant or spacetime point, the diffeomorphism is not a transformation between two different ways of describing what is happening locally. Indeed, it is a transformation between two different instants or spacetime points. In other words, diffeomorphisms are not local gauge transformations.

Above I mentioned that local observables are invariant under local gauge transformations. This motivates the technical definition of observables in phase space. However, as diffeomorphisms are not local gauge transformations, local observables are not invariant under the action of diffeomorphisms at a coordinate point. For instance, in the example of the temperature field in spacetime, I have argued that it is an observable, but the field $T(x^\mu)$ is not invariant under diffeomorphisms, as it may take different values at different spacetime points. In this sense, there is no good reason to ask for invariance under diffeomorphisms.

When analyzing diffeomorphisms from the Hamiltonian perspective, the same conclusion holds. The action of models with an explicit and dynamical diffeomorphism invariance has a local symmetry, which means that in the Hamiltonian formalism we find similar structures to the ones of gauge theories, namely, constraints and under-determined dynamics (i.e., dynamics depending on free functions $v^\alpha$). By analogy, one can build gauge generators $G_a$ associated with diffeomorphisms. The $G_a$ generate transformations according to equations \ref{eqn_gauge_transformation_1}-\ref{eqn_gauge_transformation_2}.These transformations are precisely the phase-space version of diffeomorphisms, up to the subtleties mentioned in footnote \ref{footnote_defns_gauge} . Therefore, our conclusions are translatable to the Hamiltonian setting: $G_a$ generate global gauge transformations (diffeomorphisms) which cannot be interpreted as gauge transformations in a local sense. For instance, they transform $\phi(x^\mu)$, the value of a scalar field at a point in spacetime, into $\phi'(x^\mu)$, the value that this field takes at some other point.

Finally, we can turn to the technical definition of observable (def. \ref{defn_observable_technical}). The condition \ref{eqn_invariance} used for the definition is a condition of invariance: the functions $O(t)=O(q(t),p(t))$ that satisfy the condition are invariant under the transformations generated by the $G_a$. However, as I have just discussed, for the case of diffeomorphism-invariant models, it does not make sense to ask for invariance. That is, the intuitively observable quantities (quantities satisfying def. \ref{definition_local_observable}), such as a scalar field in general relativity, are not invariant under diffeomorphisms. In other words, under a diffeomorphism $\phi (x^\mu)\neq \phi' (x^\mu)$, but it represents a perfectly well-defined, observable quantity. For this reason, the technical definition of observables in terms of Poisson brackets is inapplicable to the case of diffeomorphism-invariant models.

%
%
%

\section{Observables are not frozen or why we shouldn't change our interpretation of general relativity}\label{sect_consequences}

As I have noted above, the technical definition of observable as a quantity that satisfies the condition \ref{eqn_invariance} becomes a condition of invariance under diffeomorphisms. This implies that the invariant quantities take the same value for any value of the temporal or spatiotemporal coordinates. In other words, the only quantities that satisfy the condition are constant functions . Clearly, this goes against the interpretation of diffeomorphism-invariant models I have given above, which allowed for observables that are not constant.

Nevertheless, some authors have argued that it is possible to make sense of this family of observables and keep the original content of diffeomorphism-invariant models. These authors claim that these `frozen observables', i.e., non-evolving quantities, encode the predictions of the model. This kind of position has been defended in \citep{Anderson1995, Rovelli1990, Rovelli1991a, Rovelli1991, rovelli_gps_2002, rovelli_quantum_2004, rovelli_what_1991} and more recently in \citep{hohn_equivalence_2021, Hohn2021, dittrich_can_2017, Dittrich2015a,rovelli_philosophical_2022}. 

Furthermore, these observables would correspond with `correlations', understood as the values that some quantities take when some other quantities take some given values. For instance, in a Jacobi model for two bodies, one could define a quantity $Q_2(q_1,q_2,p_1,p_2,Q_1)$ which would represent the values that the position of the second body takes when the first one is at the position $Q_1$. For some models, this kind of construction works well and allows defining quantities that are invariant when the phase-space coordinates are transformed by a diffeomorphism.

The frozen observable view of diffeomorphism-invariant models is to take this construction and generalize it to any diffeomorphism-invariant model. `Observables' would correspond to correlations between physical quantities, and they would satisfy the invariance relation \ref{eqn_invariance}. According to this view, evolution, for diffeomorphism-invariant models, is relational, in the sense that it is encoded in these relations between variables and not represented as evolution in spacetime. Some authors in the foundations and philosophy of physics \citep{Earman2002, Earman2006, rickles_chapter_2008, rovelli_forget_2011} have taken this to have deep consequences at the time of interpreting diffeomorphism-invariant models in general and general relativity in particular.This is in contrast with the view I have developed in this article, and I believe it presents some important shortcomings. In this section I present some novel arguments discussing in depth these shortcomings.

As it should be clear from my discussion in the previous section, as I reject interpreting diffeomorphisms as local gauge transformations, I reject defining observables as invariant quantities under diffeomorphisms, i.e., I reject definition \ref{defn_observable_technical} for this kind of symmetry. Physical properties vary in spacetime and are not locally invariant under diffeomorphisms. Similarly, I reject that evolution is relational in the sense that it is encoded in correlations. That is, in our spacetime models, diffeomorphism-invariant or not, there are more predictions than just correlations, as there is a richness of spatiotemporal relations that go beyond those correlations.

There are two charitable readings that one can make of the frozen observable view. One is the defense that this position is mandated by the quantum version of diffeomorphism-invariant theories. This motivation risks being circular, as the quantization is often justified in the classical analysis. The other one is to consider phase space not as a space of instantaneous states but as a space of initial conditions. As we can find the whole evolution of the system or the properties of the whole spacetime given a set of initial conditions, then one can interpret each point in phase space to represent whole dynamical trajectories or spacetimes. Then, transformations in phase space can be interpreted as global transformations as they transform between different ways of representing whole histories of the system. This perspective is present in some works of the defenders of the frozen observables view \citep{rovelli_quantum_2004}.

Does adopting this view justify the technical definition of observable? In my opinion, it doesn't. If we study a theory like electromagnetism from the perspective that points in phase space represent the state of the electromagnetic field in the whole spacetime, we are still justified in claiming that observables are invariant under the gauge transformations of electromagnetism, and they shouldn't change if we change the initial conditions by a set of gauge-equivalent initial conditions. As the physical quantities along the evolution will depend only on the invariant quantities at the initial time (the electromagnetic field), they should be independent of the way we represent these initial conditions, i.e., which 4-potential we choose.

This motivation is not available for the case of diffeomorphism-invariant theories. That is, when we transform initial conditions using equations \ref{eqn_gauge_transformation_1}-\ref{eqn_gauge_transformation_2} for the appropriate generator, we do not obtain a set of initial conditions that represent the same instantaneous state of affairs. What we obtain is a different time slice of the evolution. Therefore, one cannot motivate the definition of observables just in the same way as one would in the case of gauge theories like electromagnetism.

The way the definition could be justified is by noting that initial conditions related by a diffeomorphism produce diffeomorphism-equivalent solutions of the equations of motion. In other words, it doesn't matter if we pick the state of the system at one moment or one second later as the initial conditions, as they will allow us to recover the same evolution. If we take it that predictions should be independent of which moment of time we pick as defining initial conditions, then we arrive at the conclusion that observables should be invariant under diffeomorphisms.

Even if this motivation seems fine at first sight, it is a bit problematic. To start with, notice that there is nothing really related to diffeomorphisms in the motivation. In other words, it could also apply to systems without an explicit diffeomorphism invariance. That is, for any system, we could demand that the predictions are made independently of which instant of time is taken as initial. This would entail demanding invariance under the transformations that shift the initial conditions, i.e., the transformations generated by the Hamiltonian. Observables, for any theory, would also be `frozen observables'.

Clearly, for theories like Newtonian mechanics, we do not follow this line of thought and we do not define frozen observables nor demand invariance under temporal translations. Still, we are able to extract all the predictions from these models and have no problem understanding them. Even if predictions are made with reference to some arbitrary instant of time, this represents no problem. For instance, a model may predict that at $t=1 s$ a body will be at position $x_1$, given that at $t=0 s$ it was at $x_0$ and had momentum $p_0$. We could express the prediction in an invariant manner in terms of a function $X_1(x_0, p_0, x_{in}, p_{in})$, i.e., a function that gives the position of the body 1 second after it was at $x_0$ with momentum $p_0$ given some initial conditions $x_{in}, p_{in}$. This would be a `frozen observable', as it wouldn't change if we changed the initial conditions $x_{in}, p_{in}$ to other initial conditions $x'_{in}, p'_{in}$ along the same trajectory. 

A moment of reflection suffices to reach the conclusion that this construction is only a contrived way of expressing what we already knew from the analysis of the original model. In this sense, making some reference to a reference point, or using some coordinates, does not represent any problem for our understanding of the model. In the case of diffeomorphism-invariant models, including general relativity, the same holds. For instance, in the example of the rocket, we can use a coordinate system to predict the value of the temperature field at the spacetime point which lies in the trajectory of the rocket and is one hour into the future of its taking off. This is a perfectly fine prediction, and we wouldn't gain any knowledge or information by expressing it as a complicated function of the initial conditions. In this sense, there is no need to push for this definition of observables in terms of constants under different choices of initial conditions on a trajectory.

Finally, even if one wanted to insist on the frozen observable view, the claim that it proves that evolution is relational or that all the predictions of the theory are correlations between observables is false. That is, for sure correlations can be expressed as invariant functions of the initial conditions, but the rest of the content and predictions of the model would still be there. For instance, if we followed the correlation view, the predictions of the Jacobi model of $n$-bodies, i.e., the diffeomorphism-invariant version of a model describing the dynamics of a system of $n$ bodies, would correspond to things like the position of body 1 when bodies 2 and 3 are at some given positions. But surely the model makes other predictions, as it not only predicts that some configurations will obtain, but also the order in which they obtain. Similarly, it predicts the time that elapses between two configurations, even if it cannot be read directly from the arbitrary temporal parametrization of the model. This could be written as a phase space function of the form $T(Q^1,Q^2,q_i,p_i)$, where $Q^1$ and $Q^2$ are the two configurations and $q_i,p_i$ are a set of initial conditions. Changing these initial conditions along a trajectory wouldn't change the value of the function.

In this sense, the frozen observable view falls short and misses much of the structure of the theory. What we would be left with is not a reduced set of quantities but exactly the same quantities and predictions as we had before. This was to be expected, as sets of initial conditions represent whole solutions, with all their structure. Now, given all this reformulation in terms of constant quantities along trajectories, does anything change about what we took to be the empirical content of the diffeomorphism-invariant model? To me, it is clear that it doesn't. That is, even if rewritten in terms of complicated functions, our models still describe spacetimes composed of spacetime points in which different fields and bodies take some configurations and the relations between these points. The observable quantities, in the intuitive senses defined in definitions \ref{definition_local_observable} and \ref{definition_global_observable} remain the same.

To insist, there is no problem in understanding a point in phase space as representing an instantaneous configuration of a system and to understand its phase space coordinates as local observables in the sense of definition \ref{definition_local_observable}. From the global perspective that interprets phase space points as representing whole trajectories, it is fine (even if there is no necessity) to write predictions in terms of invariant quantities, but it does not change the way our models should be interpreted. 

Let me make a technical remark. I have argued that any prediction of a deterministic theory can be rewritten as a function of a set of initial conditions and that one can build it so that it remains constant if the set of initial conditions changes according to the dynamics of the theory. However, it is important to note that these functions may be very complicated, and, in particular, may not be continuous and differentiable functions in phase space. They may even be defined only on some subsets of phase space. When this happens, one cannot define Poisson brackets for these quantities, and the technical definition of observable (def. \ref{defn_observable_technical}) does not apply in this case. This would motivate a refinement of the definition, although it would complicate the quantization of diffeomorphism-invariant models following the frozen observable perspective, as some of the defenders of this view have acknowledged \citep{dittrich_can_2017, Dittrich2015a, chataignier_relational_2024}.

\begin{figure}[ht]
\centering
\includegraphics[alt={Three side-by-side plots showing 2D trajectories in a plane, with the first plot's horizontal axis labeled q1 and vertical axis labeled q2. From left to right, the plots show increasingly complex curves bounded within a fixed rectangular region. The left plot features a simple, closed, elongated loop representing a basic periodic orbit. The middle plot shows a more intricate, overlapping closed Lissajous-style curve. The right plot displays an extremely dense, continuously overlapping curve that blackens almost the entirely of the rectangular area, illustrating a space-filling, quasi-periodic, or chaotic trajectory.},width=0.9\textwidth]{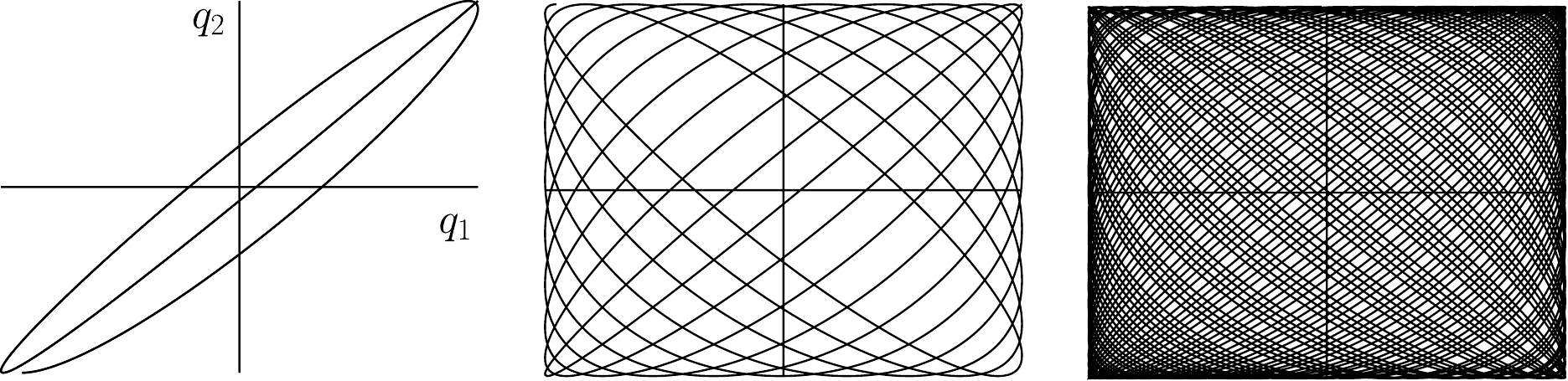}
\caption{\label{figure_oscillators} Configuration space trajectories of two harmonic oscillators}
\end{figure}

These points will become clearer with an example. Consider a temporal reparametrization invariant model that describes a pair of harmonic oscillators with non-commensurable frequencies, that is, $\frac{\omega_1}{\omega_2} \notin \mathbb{Q}$. The trajectory in configuration space of this system is quasi-periodic and dense in a region, as represented in figure \ref{figure_oscillators}.

A characteristic feature of this system is that if it goes through a point in phase space, it only does it once, so it can be used for defining correlations with a third system. For instance, we could define the correlation $Q_{3,0,0,+}(q_1,p_1,q_2,p_2,q_3,p_3)$ that would encode the configuration of the third system when both  oscillators are at position $0$ with positive momentum. However, only a measure-zero set of trajectories ever passes through this point, so the function is defined only along those trajectories. This makes it non-differentiable. 

Similarly, we could define functions encoding temporal relations for this system. Let $x$ denote a point in the phase space of the two oscillators, then we can define a  time ordering function $TO(x,q_1,p_1,q_2,p_2)$ which takes the value $1$ if it $x$ happens after the oscillators travel through the origin (with positive momenta) and $-1$ if it happens before. As before, this function is defined only if $q_1,p_1,q_2,p_2$ and $x$ are part of a same trajectory that at some moment crosses the origin.

Both functions exemplify what I was arguing before: any prediction of the model can be expressed as a (complicated, non-continuous) function of a point in phase space that can be taken as initial conditions. These functions of course remain constant if we move along dynamical trajectories. Again, this is just a consequence of the fact that the predictions of a deterministic model do not depend on which point on a dynamical trajectory is chosen for making the predictions.

Similarly, this example illustrates well the difference between observables in the intuitive sense and observables understood as any phase space function satisfying an invariance condition. The functions I have just defined satisfy an invariance condition and it is true that we are able to compute them if we know the state of the system at some moment, but that does not make them observable in the most direct sense of observable. That is, by observing the two oscillators now and knowing the equations of motion of the system, I am able to work out whether the two oscillators will have position $0$ simultaneously and, if this happens, compute the position of a third system at that time. Inferring a future position is conceptually different from observing it.

The bottom line of this example is that even if in diffeomorphism invariant models we can write the predictions of the model as phase space functions satisfying certain invariance conditions, this does not change the empirical content of the model nor the way we ought to interpret it. The system of two harmonic oscillators is best understood as a system of two harmonic oscillators that evolves in time. Local observables, i.e., properties of the system at a time, are the positions of the oscillators, their velocities, and some simple functions of them (e.g., their energies or angular momenta). Global observables include predictions about whether certain configurations obtain, the order in which they do, and the Newtonian time that elapses in between configurations.  

Another point that I think is worth raising is that in the literature one can find claims such that to find the empirical content of a diffeomorphism-invariant model one needs to find its observables in its technical sense. On top of all the objections above, I want to add a further problem with this sort of claim: some of the `observables' won't have anything to do with the model. 

Consider the Jacobi action for a system of two free bodies. As the constraint that generates diffeomorphisms in this model is $C=\frac{p_1^2}{2m_1}+ \frac{p_2^2}{2m_2}$, one can readily check that $p_1$, $p_2$, and $q_2-\frac{m_1p_2}{m_2p_1}q_1$ are constants of motion and, therefore, invariant under diffeomorphisms. With these quantities, one can build the `correlation':
\begin{equation}
Q_2(Q_1,q_1,q_2,p_1,p_2)=\frac{m_1p_2}{m_2p_1} Q_1 + (q_2-\frac{m_1p_2}{m_2p_1}q_1) \, ,
\end{equation}
which is a constant of motion that represents the position of the second body when the first one is at position $Q_1$. This is the standard frozen observable authors in this literature speak about. 

But nothing prevents us from building also the following quantity:
\begin{equation}
\tilde{Q}_2(Q_1,q_1,q_2,p_1,p_2)=\frac{17p_2}{13p_1} Q_1^2 + 5 Q_1+ (q_2-\frac{m_1p_2}{m_2p_1}q_1) \, ,
\end{equation}
this quantity is also a constant of motion (it is built from constant functions), and it has nothing to do with the correlations that this system will exemplify. Indeed, it can be interpreted as a correlation, but not of the system of two particles, but of a model in which particle 1 is free, while particle 2 is uniformly accelerated. In this particular expression, I have chosen the initial velocity of $Q_2$ (with respect to $Q_1$) to be $5$ and the acceleration to be given by $\frac{34p_2}{13p_1}$. These choices were completely arbitrary to show how there is a huge freedom in building frozen observables that have nothing to do with the actual system. We could have set the acceleration to be the gravitational acceleration, and we could have built a correlation for far more complicated and equally fictitious dynamics.



Both $Q_2$ and $\tilde{Q}_2$ are `frozen observables'. Both represent more or less complicated phase space functions that can be computed from the state of the system at a time to make an inference about the future of the system or about another hypothetical system. None of them therefore seems to be observable in the local intuitive sense: while it is uncontroversial to claim that we observe $q_1,q_2$, it seems far more contestable to claim that we observe $Q_2$ and $\tilde{Q}_2$. From the point of view of the local definition of observable (def. \ref{definition_local_observable}), this kind of argument should make us reject frozen observables as local observables at a moment represented by a phase space point. 

To make the point more vivid, imagine a situation in which you invite someone to observe the two-body system at some moment of time. If you asked this person what they have observed, they would probably tell you something equivalent to $q_1,p_1,q_2,p_2$. Then it clearly seems absurd to ask, `haven't you \textit{observed} what the position of the second body will be when the first one is at $Q_1$?'. The poor observer clearly does not have the resources to answer this question. If we don't tell them that the dynamics is such and such, they cannot tell us that this position will be $Q_2$ and not $\tilde{Q}_2$ or some other arbitrary value. Even if we explained in full detail the dynamics of the system to them so that they can make the right prediction, it would go clearly against the common usage of the term `observe' to claim that they have observed $Q_2$.

But even if we consider a global perspective were $Q_2$ and $\tilde{Q}_2$ are understood as functions in a space of initial trajectories, the fact that $Q_2$ represents a correlation that will obtain for the actual system, while $\tilde{Q}_2$ represents a correlation of an hypothetical system that has nothing to do with the system, represents a challenge for the frozen observable doctrine. 

Now imagine we asked again our observer friend to observe the system, but for the whole evolution of the system, not just an instant as before. After this extended in time observation, it would make sense to ask the observer what the position of the second body was when the first one was at $Q_1$. We know that this observation corresponds to $Q_2$ and not $\tilde{Q}_2$. As in the instantaneous case, we could ask the observer what the observation would have been had the dynamics of the second body been uniformly accelerated under certain conditions. The observer could do the maths and give us the right answer, but it is hard to imagine that they would claim that they have observed $\tilde{Q}_2$.

This kind of example shows how the definition of frozen observable needs to incorporate some restriction in order to distinguish between actual predictions of a model and spurious quantities that satisfy the invariance condition. What kind of restriction could do this work? The predictions of diffeomorphism-invariant quantities are clear. We will have predictions like correlations, but also predictions like spatiotemporal relations. All of them can be written as invariant quantities (again, if the differentiability condition is relaxed). Unfortunately for the recipe-seeking physicist, the way of telling apart actual predictions, from arbitrary invariant functions, requires us to look at the models and interpret them, just as for telling apart $Q_2$ from $\tilde{Q}_2$ we had to look at both quantities and identify one of them as corresponding to a real prediction and the other one as corresponding to a fictitious system. In this sense, it is not that we start with a formalism, look for some quantities satisfying certain conditions, and then give an interpretation for it. The order is reversed. We start with a  theory for which we have a satisfactory interpretation. We know what predictions can be made from a theory like general relativity. And now, if we want, we can build (highly-complicated) functions in phase space that encode these predictions. Again, this is not necessary nor it illuminates us in any manner that forces us to change our interpretation of the theory.

To sum up, the grain of truth of the frozen observable view is that the predictions of a diffeomorphism-invariant model can be written as phase space functions that are invariant under diffeomorphisms. However, this does not mean that we ought to change our interpretation of diffeomorphism-invariant models. These models describe spacetimes (or systems evolving in time), the distribution and evolution of some degrees of freedom, and the spatiotemporal properties and relations of different parts of this spacetime. Therefore, we can still think about observables in this kind of model as I have argued above. We know the local and global observables of this kind of model. The technical definition of observable in phase space (defn. \ref{defn_observable_technical}), when modified to acknowledge for non-differentiable invariant quantities, captures the fact that predictions can be written in an invariant manner, but fails to give local observables in the intuitive sense of def. \ref{definition_local_observable}.

\section{Conclusion}\label{sect_conclusion}

In this article I have analyzed how the concept of observable from gauge theory has been applied to the case of diffeomorphism-invariant models, most notably, general relativity. I have distinguished between two notions of gauge transformation, local and global, and I have argued that diffeomorphisms can be conceptualized as gauge transformations only as global transformations. The local action of a diffeomorphism is not to change the way we represent the state of affairs at a point in spacetime or instant of time, but to shift the spacetime point or instant of time represented by a point in a manifold or set of coordinates. In this sense, I have argued that it does not make sense to ask for local invariance under diffeomorphisms or to define observables as invariant quantities. Diffeomorphism-invariant models, just as any other non-quantum model in physics, describe what happens to matter and field degrees of freedom in a set of instants or spacetime points, as well as the spatiotemporal relations between the different events.

This position has consequences for the foundations of general relativity and quantum gravity. More precisely, the analysis in this article leads one to reject the claim that the gauge analysis of diffeomorphisms implies that the physical content of diffeomorphism-invariant models lies in correlations and the claim that our understanding of space and time should be radically different in this sort of theory. Similarly, it can be the basis for arguing that `timeless' theories of quantum gravity are based on misguided analysis of diffeomorphisms as gauge symmetries and that this should raise concerns about these theories.

\section*{Acknowledgements}

I am grateful to two anonymous reviewers for their very helpful and constructive comments. I also want to thank Brian Pitts and Carl Hoefer for discussions over the years on the topic.


\begin{flushright}
\emph{Department of Architectural Technology, Division of Mathematics\\
Universitat Politècnica de Catalunya\\
Barcelona, Spain\\
alvaro.mozota@upc.edu\\}
\end{flushright}

\bibliographystyle{chicago}
\bibliography{References.bib}

\end{document}